\documentclass[authoryear,12pt]{elsarticle}

\usepackage{amssymb}

\journal{NIM}

\begin{document}
\begin{frontmatter}



\title{GPS Timing and Control System of the HAWC Detector.}


\author[MSUAuthers,UU]{A. U. Abeysekara$^*$}
\author[MSUAuthers]{T. N. Ukwatta}
\author[MSUAuthers]{D. Edmunds}
\author[MSUAuthers]{J. T. Linnemann}
\author[UW]{A. Imran}
\author[LANL]{G. J. Kunde} 
\author[UW]{I. G. Wisher}

\cortext[corresponding]{Corresponding author: udaraabeysekara@yahoo.com}

\address[MSUAuthers]{Department of Physics and Astronomy, Michigan State
University, East Lansing, MI, USA}
\address[UU]{Department of Physics and Astronomy, University of Utah, Salt Lake City, UT, USA}
\address[UW]{Wisconsin IceCube Particle Astrophysics Center (WIPAC) and Department of Physics, University of Wisconsin-Madison, Madison, WI, USA}
\address[LANL]{Physics Division, Los Alamos National Laboratory, Los Alamos, NM, USA}

\begin{abstract}
The design and performance of the GPS Timing and Control (GTC) System of the High Altitude Water Cerenkov (HAWC) gamma ray observatory is described.
The GTC system provides a GPS synchronized absolute timestamp, with an accuracy better than 1$\mu$s, for each recorded event in HAWC.
In order to avoid any slack between the recorded data and the timestamp, timestamps are injected to the main data acquisition (DAQ) system 
after the Front-end Electronic Boards (FEBs).
When HAWC is completed, the HAWC main DAQ will use 10 time to digital converters (TDCs).
In order to keep all the TDCs in sync, the GTC system provides a synchronized clock signal, coordinated trigger signal, and control signals to all TDCs.

\end{abstract}

\begin{keyword}

GPS timestamp \sep gamma-ray astrophysics \sep water cherenkov detector
\sep time to digital converter \sep TeV astronomy
\end{keyword}

\end{frontmatter}


\tableofcontents

\section{Introduction}\label{Introduction}
The GPS Timing and Control (GTC) system is one of the major subsystems in the High Altitude Water Cherenkov (HAWC) gamma ray observatory \citep{HAWCGRB, HAWCConstruction}.
HAWC is a very high energy (VHE) gamma ray observatory being built on the flank of the volcano Sierra Negra in
Mexico at latitude $18^{0} 59' 48''$ North, longitude $97^{0} 18' 34''$ West, and altitude 4100 m.
It is important for HAWC to maintain a low dead time, as an all-sky survey instrument.
In order to maintain a low dead time, the HAWC main data acquisition system (DAQ) was designed as a distributed DAQ, providing continuous read out.
The first level of the DAQ consists of 11 Caen VX1190A Time to Digital Converters (TDCs) and 11 
GE XVB602 Intel Corei7 Single Board Computers (SBCs) to read out TDCs, 
where each TDC-SBC pair reads a fragment of an event with 128 channels.
These event fragments are combined in the online reconstruction farm.
The distributed design of the DAQ system makes synchronous operation of the system critical.
The GTC system, which is a custom FPGA based system, is designed to perform this task.\\

As its name suggests, the GTC system has two sub systems: the GPS Timing system and the Control system.
The primary task of the GPS Timing system is to provide a timestamp for each recorded event, which is the absolute time of the trigger.
The primary task of the Control system is to provide the clock and the control signals to the TDCs, and provide trigger and  detector status information to the scaler system.
The GTC system is implemented using three different types of custom cards: Clock type HClock Card, Control type HClock Card and CB\_Fan Card, as well as a commercial GPS receiver NAVSYNC CW46S \footnote{http://www.navsync.com/docs/cw46s$\_$pb.pdf}.\\

\section{HAWC Observatory}

The completed HAWC detector will consist of 300 steel water tanks of 7.3 m in diameter and 4.5 m in height instrumented with 4 PMTs in each tank.
Each of these tanks contain a light-tight bladder filled with purified water and 4 PMTs pointed upwards are placed near 
the bottom of the bladder.
Construction of HAWC is scheduled in stages; continuous operation of the first phase with 30 tanks (HAWC 30) with a fully functional GTC system started in November 2012, and the final phase with 300 tanks will be completed in 2014.
\\

The HAWC detector is designed to observe cosmic gamma rays by detecting the component of Extensive Air Showers (EAS) which reaches ground level. EAS are generated from the interactions between the earth's atmosphere and cosmic gamma rays.
When the relativistic charged particles in an EAS move through the water tanks, they create Cherenkov light that can be detected by the PMTs.
The main DAQ measures the arrival time and Time Over Threshold (TOT) of the PMT pulses, with an accuracy of 100 ps, using Caen VX1190A Time to Digital Converters (TDCs) \footnote{http://www.caen.it}.
This information is used to determine the species of the primary particle initiating the EAS (gamma ray or proton), its energy, and the celestial coordinates of the primary particle.\\

\section{Data Acquisition System (DAQ)}

Caen VX1190A TDCs are designed to record TOT measurements within a given time window, around a trigger signal.
Each of these TDCs is equipped with 128 data channels and an output buffer to store data until read out.
The control of the TDCs is done using three signals of the TDC control bus: TRG, CLR, and CRST.
The TRG is the trigger signal input to the TDC.
In HAWC, the trigger signal is a periodic signal that is provided by the GTC system.
In a typical data run the periodic trigger frequency is 40 kHz (period = 25 $\mu$s) and the TDCs record the data in a 25.2 $\mu$s window around each trigger.
The data saved in a given time window is called an ``event" in this paper.
The analysis software searches these events for individual EAS arrivals.  The overlapping of the ``event'' time windows and the ability of the TDCs to read out while acquiring new data provides read out with no intrinsic DAQ dead time.
CLR is the clear command, which clears the data in the output buffer, resets the event counter, 
bunch counter\footnote{Bunch counter is a 12 bit counter that counts number of 40 MHz clock cycles from 
the last bunc counter reset.}, and performs a TDC global reset.
CRST is the reset command, which resets the extended trigger time tag and bunch counter.\\

\begin{figure}[h!]
\centering
\includegraphics[width=\textwidth]{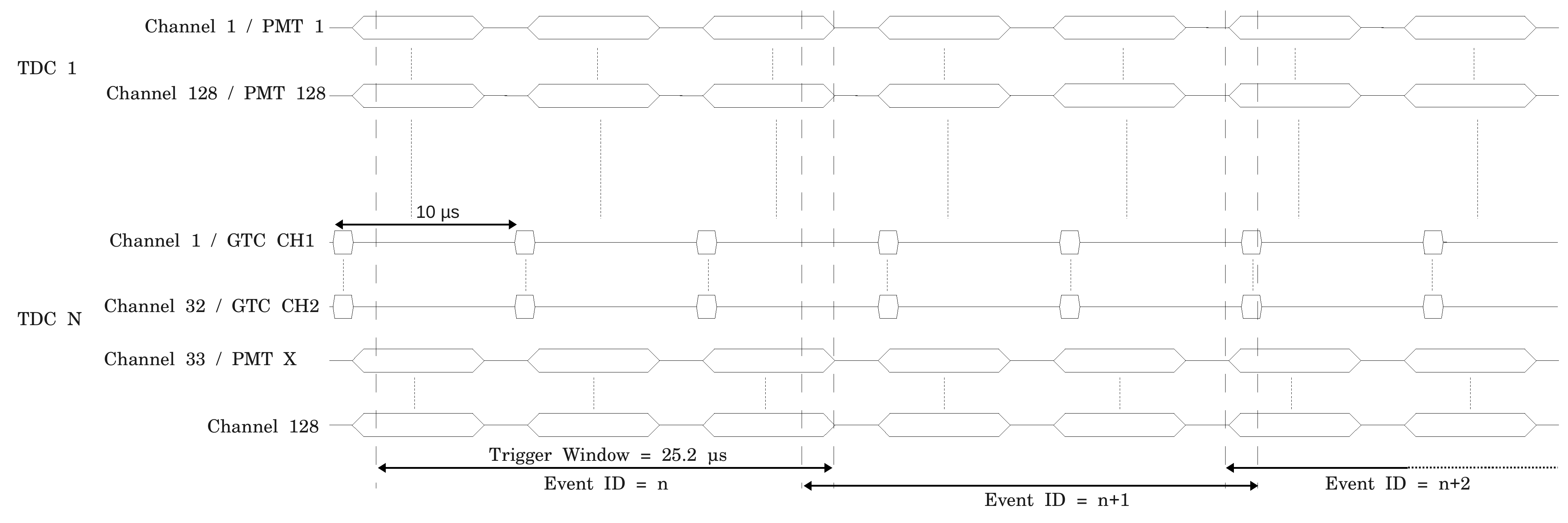}
\caption{A simplified timing diagram of the PMT signals and the timestamp signals are shown.
Channels 1 through 128 of TDCs 1 through (N-1) record the PMT signals and Channels 1 through 32 of the N$^{\textrm{th}}$ TDC record timestamps.
Two timestamps per each trigger window are guaranteed, when the Clock System is configured to send a timestamp in every 10 $\mu$s.}\label{OVerallTiming}
\end{figure}

When HAWC is completed, it will need 1200 data channels\footnote{Completed HAWC is 300 tanks, with 4 PMTs per each tank.}, which is 9 full TDCs and 48 data channels from a 10$^{\textrm{th}}$ TDC.
Besides the 10 TDCs used to record PMT signals, an $11^\textrm{th}$ TDC will be used to record 32 signals coming from the GPS Timing System and calibration signals.
These signals are similar to the TOT signals but they are encoded with the current GPS time, which is the  timestamp of that event.
Figure~\ref{OVerallTiming} shows a simplified timing diagram of the PMT signals and the timestamp signals.
In this timing diagram, Channels 1 through 128 of TDC 1 through (N-1) record PMT signals and Channels 1 through 32 of the N$^{\textrm{th}}$ TDC record timestamps.\\

While TDC buffers are filling with data, a GE XVB602 Intel Core i7 based VME Single Board Computer (SBC) reads each TDC and delivers the data to the online reconstruction farm.
However, SBCs cannot perform the read out process at exactly the same rate for every TDC.
Therefore, the online reconstruction farm receives different fragments of a single event at different times.
The HAWC online reconstruction software identifies the event fragments belonging to a given trigger using the event identification number (Event ID), which is a 12 bit number in the event header. 
The Event ID becomes zero after a TDC power cycle and then increases by one for each trigger.
The GTC system also can reset the Event ID to zero by sending a CLR signal through the TDC control bus.\\

After identifying the event fragments of a single event, the online reconstruction process combines the fragments into a single event and decodes the timestamp.
This event build is possible only if all the TDCs are working synchronously and maintain a unique Event ID for a given trigger.
The main objective of the Control System is to keep the TDCs in synch.
The synchronization between TDCs is achieved by distributing a global clock signal to all the TDCs, and clearing and resetting all the TDCs simultaneously at the beginning of each run.

\section{The GPS Timing System.}

The GPS Timing System provides two services to HAWC: 1) produce a periodic timestamp and 2) derive a low jitter 40 MHz signal to use as the global clock signal for HAWC.

\subsection{Hardware implementation.}

\begin{figure}[h!]
\centering
\includegraphics[width=\textwidth]{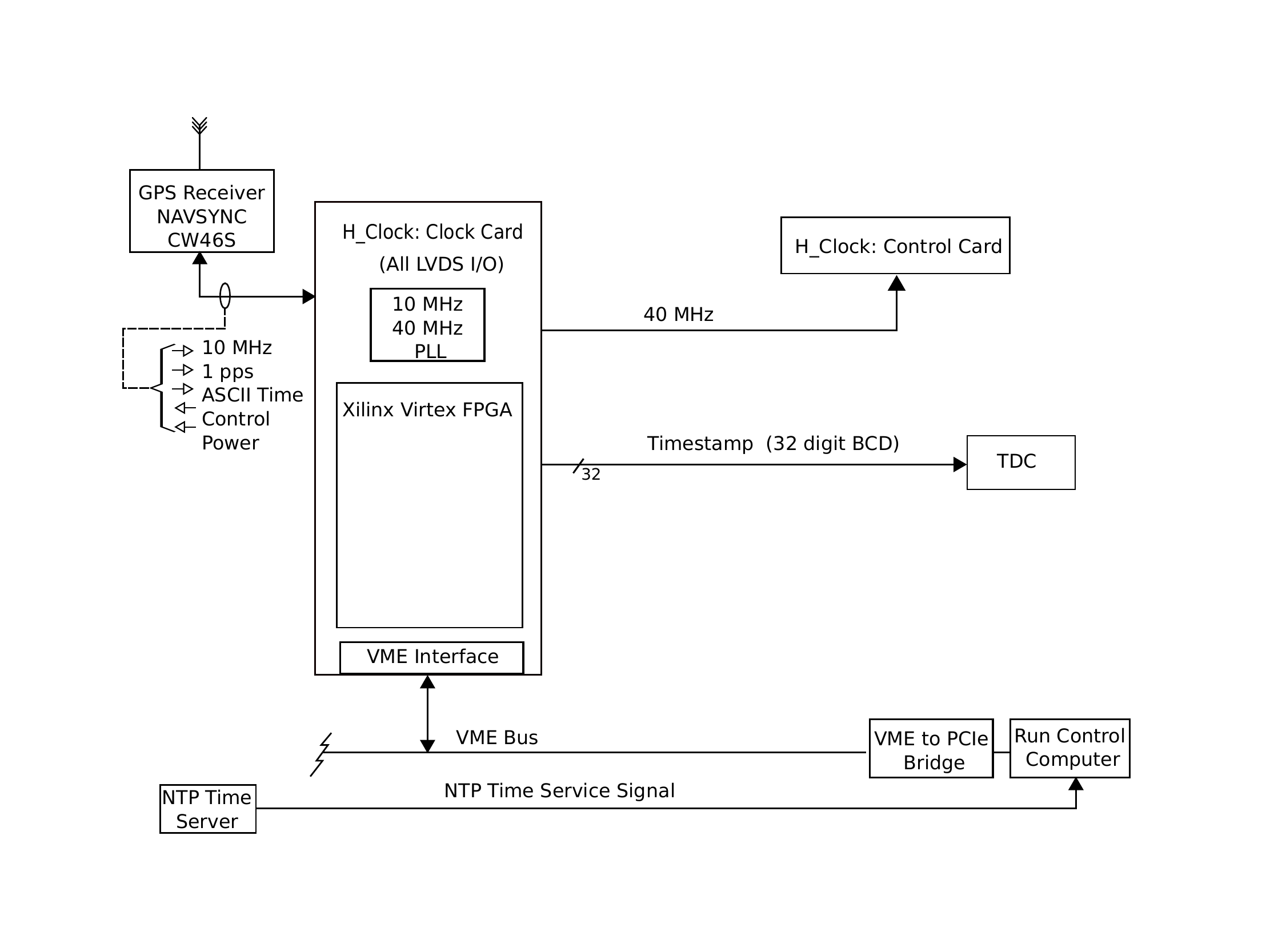}
\caption{A block diagram of the Clock System is shown.
         The NAVSYNC CW46 GPS receiver and the Clock Card are the two main components of the Clock system.
         The GPS receiver is used to obtain the GPS time and a 10 MHz signal.
         The Clock Card produces a 40 MHz global clock signal and two timestamps for the TDC and scaler systems.
         The communication between the Clock Card and the control computer is done through a VME A24D16 interface.
}\label{ClockCardSchem}
\end{figure}

\begin{figure}[h!]
\centering
\includegraphics[width=100mm]{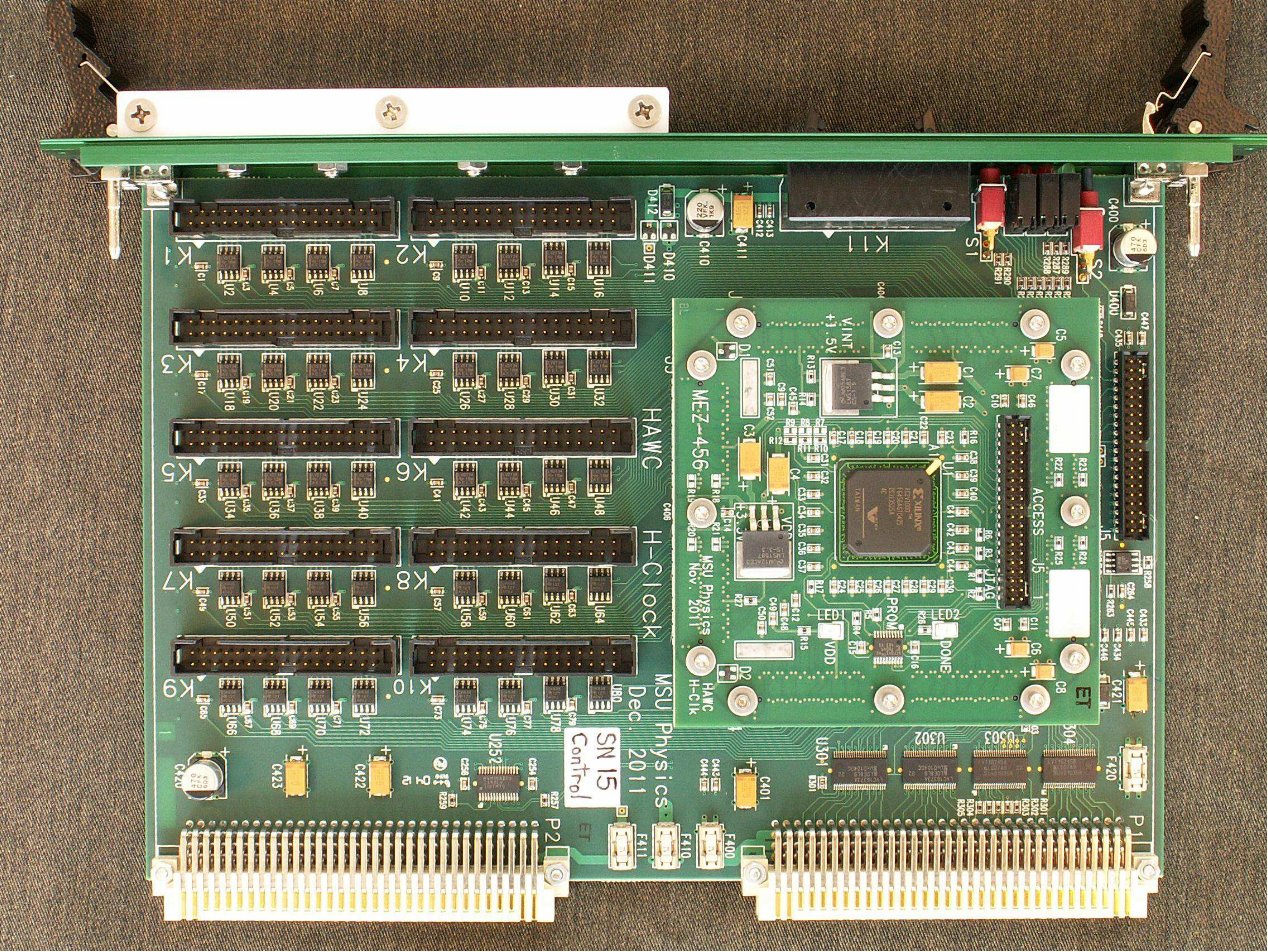}
\caption{A photograph of a fully assembled HClock card is shown.
         The version of the HClock card that is used in the Clock System has 2 general purpose input ports and 8 general purpose output ports.
         The version of the HClock card used in the Control System has 4 general purpose input ports and 6 general purpose output ports.}\label{HclockCard}
\end{figure}

As shown in Figure~\ref{ClockCardSchem}, the GPS Timing System is made from two components: a custom board called the Clock type HClock Card and a NAVSYNC CW46S GPS receiver.
Figure~\ref{HclockCard} shows a photograph of a fully assembled HClock card
\footnote{Note that a Clock type HClock Card is a version of the HClock card.}.
It is a 2 slots wide 6U VME-64X module that is equipped with a Phase Lock Loop (PLL), ten 17-pair (34 pins) LVDS General Purpose Input Output (GPIO) ports, a 16 pin connector to the GPS receiver, a A24D16 VME interface and  a Virtex II FPGA.\\

Each of these GPIO ports has 16 LVDS GPIO signals to/from the FPGA and the $17^{th}$ pair carries a 40 MHz clock signal, which is also an LVDS signal.
The direction of the GPIO ports is switchable by changing the IO driver chips.
Clock type HClock cards are made with two input ports and  eight output ports.\\

The FPGA is mounted in a mezzanine card (labeled as Mez-456 in the picture).
Since the performance and the resources of the Virtex II family FPGAs are adequate for the requirements of HAWC, a Virtex II xc2v1000-4fg456 FPGAs in our shop's spare stock was used.
However, if a future upgrade needs to change the FPGA, it can easily be done by simply designing a new mezzanine card.\\

The GPS receiver is used to obtain the GPS time and a 10 MHz sine wave signal.
The internal PLL of the Clock type HClock Card uses this 10 MHz sine wave signal to derive a low jitter 40 MHz digital clock signal and makes several exact copies that are delivered to the Control type HClock Card, to the FPGA and to the $17^{th}$ signal pair of all the GPIO connectors.
This 40 MHz signal is used as the global clock signal of HAWC.
Other than this sine wave, the GPS receiver transmits a one pulse per second (1PPS) pulse stream and a set of data strings via the RS232 protocol.
The rising edges of these 1PPS pulses mark the top of each second.
The firmware running inside the FPGA uses this 1PPS signal and the data strings to replicate the current GPS time.\\

\subsection{Firmware Implementation.}\label{ClockFirmware}

\begin{figure}[h!]
\centering
\includegraphics[width=\textwidth]{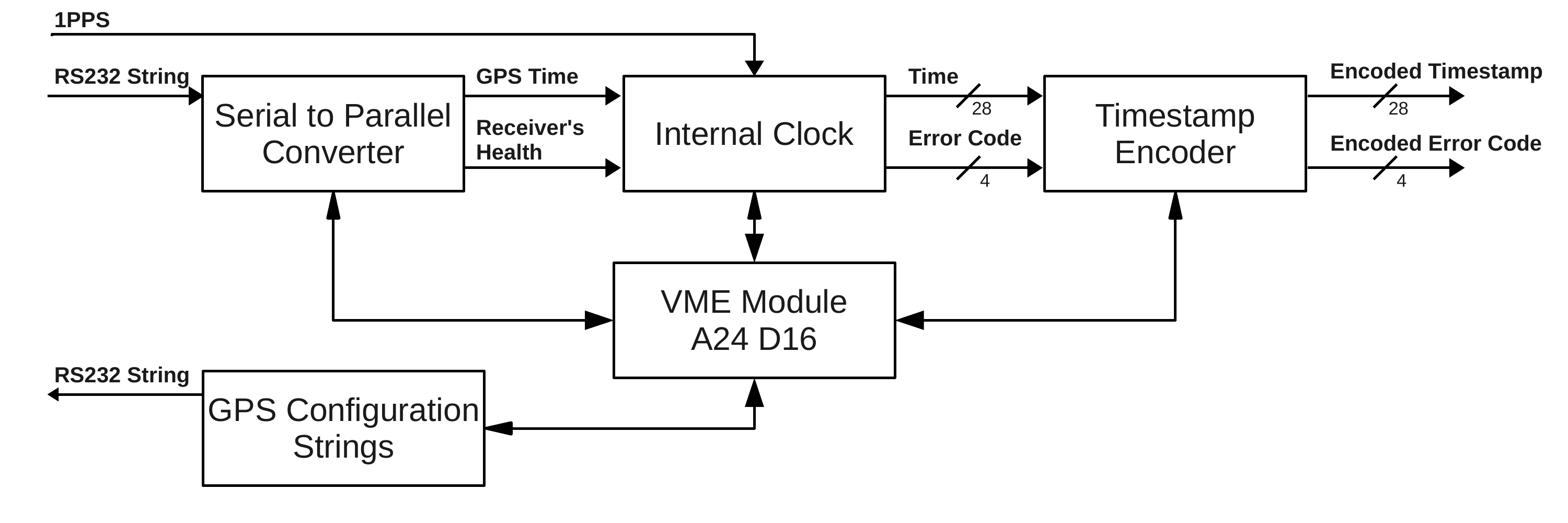}
\caption{A simplified functional block diagram of the Clock firmware is shown.
         This firmware maintains an internal clock synchronized with the GPS time and produces TDC readable timestamps.
         The communication between this module and the control computer is done by an A24D16 VME interface.
         }\label{ClockFirmwareBlockDiagram}
\end{figure}

A simplified functional block diagram of the Clock firmware is shown in Figure~\ref{ClockFirmwareBlockDiagram}.
This is a sequential logic design with several state machines implemented using VHDL.\\

The GPS receiver installed at the HAWC site is configured to send three data strings followed by a 1PPS pulse.
These three strings (POLYT, GPGSA, and POLYP) are standard NMEA 0183 strings, which carry the current GPS time, GPS receiver operating mode, number of 
visible satellites, and Dilution Of Precision (DOP) values.
The first module of the firmware reads these serial strings and extracts the current GPS time and the health information such as the GPS fix status\footnote{ GPS Fix is an integer between 1 and 3. If the GPS receiver is not getting enough GPS signals and it is unable to fix, GPS Fix is 1.
If the GPS receiver is able to make a 2D or 3D fix this GPS Fix becomes 2 or 3 respectively.
Refer to CW25 GPS Receiver User Manual for more information, see http://www.navsync.com/GPS\_integrated.html.} and dilution of precision.
Then the GPS time and health information goes to the internal clock module, which is a continuously running 8 digit binary coded decimal (BCD) clock using the 40 MHz clock signal as the reference frequency.
In this 8 digit clock the least significant digit is microseconds and the most significant digit is tens of seconds.
This clock module also receives the 1PPS signal, which is used to identify the top of each second.
At the top of each second, the Internal Clock module compares its clock time with the GPS clock time, and overwrites the internal clock if the times do not match and the GPS receiver is in good health.
This allows the Timing System to have an internal clock that runs synchronously with the GPS clock.
The final stage of the firmware is to make a TDC readable timestamp in every $\Delta T \mu$s interval, where $\Delta T \mu$s can be configured to $10~\mu$s 
or $20~\mu$s.
Other than these major modules, the Clock Firmware has a VME module that handles an A24D16 VME interface and several FIFOs 
that are filled with GPS health monitoring information.

\subsection{The Timestamp Encoding Algorithm.}\label{TSAlgo}

\begin{figure}[h!]
\centering
\includegraphics[width=\textwidth]{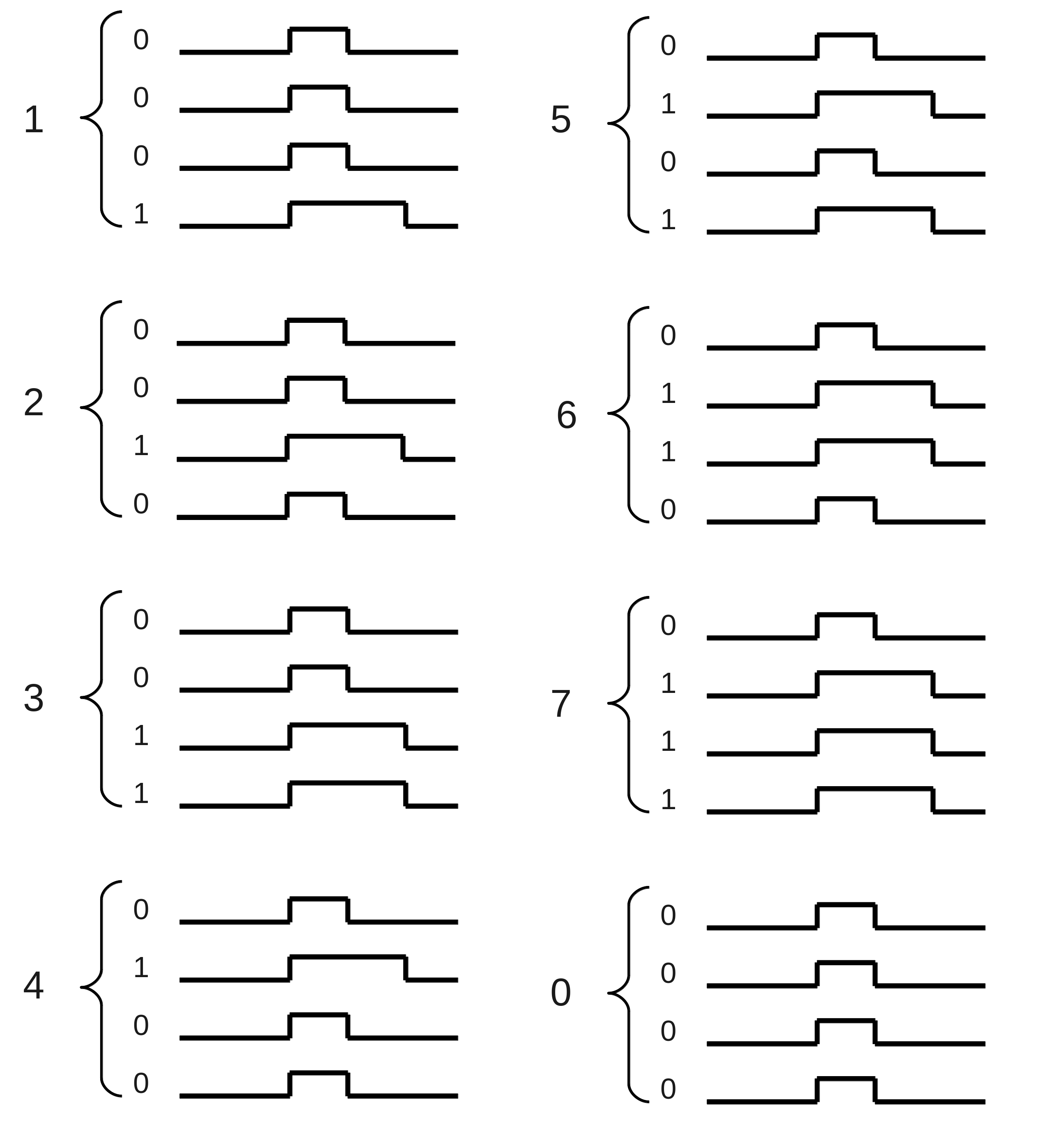}
\caption{Encoding of the timestamp 12.34567 seconds with no errors is shown.
         Each digit is encoded into a 4 bit binary number, and each binary number is encoded into a pulse.
         A 1$\mu$s wide pulse is used to indicate logic 0, and a 2$\mu$s wide pulse is used to indicate logic 1.}\label{ExampleTimeStamp}
\end{figure}

The first 28 bits of the GTC timestamp are a 7 digit BCD value that carries the time in the format:
 10s of s, 1s of s, 100s of milliseconds, 10s of milliseconds, 1s of milliseconds, 100s of microseconds and 10s of microseconds (ss:mmm:uu).
The remaining 4 bits are used to encode various
errors that are encountered in the process of acquiring and
encoding the timestamp. These error codes are defined in
Table 1.
The encoding of this timestamp to a TDC readable format is done using a simple algorithm.
Each bit is denoted by a pulse; if a pulse is 1 $\mu$s wide it denotes a logic zero bit, if a pulse is 2 $\mu$s wide it denotes a logic one bit.
As an example, the timing diagram shown in Figure~\ref{ExampleTimeStamp} is the encoding for the time 12.34567 seconds with no errors.
An encoding scheme of this type with pulses must be used because the TDCs are only sensitive to edges but not to logic levels.
That is one cannot just send the 28 raw binary bits with logic levels to the TDCs because most of the time, most of the lines will not make a logic transition during a trigger window ($25.2~\mu s$).

\begin{table}
\begin{tabular}{| c | l |}
\hline
Error Code & Error \\
\hline
xxx1 & Internal Clock was over written by the current GPS clock.\\
xx1x & GTC system lost communication with the GPS receiver.\\
x1xx & Not enough satellites to make a GPS Fix.\\
1xxx & NMEA string coming from the GPS receiver has errors.\\
\hline
\end{tabular}
\caption{The errors corresponds to each timestamp is encoded into the 4 most significant bits of the timestamp. The meaning of each error code is shown.}
\label{ErrorCodeTable}
\end{table}

\subsection{Timestamp Decoding} \label{TSDecAlgo}

Every HAWC event has a timestamp associated with it. 
This timestamp is constructed by combining three components: TDC timestamp, NTP timestamp, and trigger derived timestamp.

\subsubsection{TDC Timestamp} \label{tdc_time}
TDC Timestamp is the encoded timestamp that comes from the GTC system.
These timestamps are sent to a TDC when the microseconds digit of the absolute clock time is 0, for example when the absolute time is **.****00 sec, **.****10 sec, **.****20 sec, etc.
The TDC that records this encoded timestamp is also read out in the same way as the other TDCs, and the time stamp becomes a part of the main data stream.

\subsubsection{NTP Timestamp}\label{ntp_time}

The TDC timestamp (ss:mmm:uu) rolls over every minute.
Hence, the low-resolution timestamp (\texttt{yyyy:mm:dd:hh:mm:ss}) needs to be combined with the raw HAWC TDC data within one minute of the trigger time. 
This is done by the single board computers that read TDCs, which record the system time each time TDC readout is completed for a block of events.
The computer system clock of the single board computers is synchronized via NTP time service to a local NTP time server \footnote{http://www.ntp.org/}.
With the local NTP time server, the absolute accuracy of the system clocks is in the millisecond range.

The TDC timestamp, and the NTP timestamp of each event get combined together in the online reconstruction farm.
When they are added, the tens of seconds and the seconds digits are coming from both the TDC timestamp, and the NTP timestamp.
These two digits must be equal if both timestamps, TDC timestamp, and NTP timestamp, are accurate.
Therefore, we use this property as a sanity check to measure the accuracy of the timestamps.

As discussed before, the latency between an event trigger and the DAQ computer adding its time stamp to that event's data plus the uncertainty of the
computer's system clock should together be less than one minute when the DAQ system is running normally. In the current configuration, the SBCs initiate readout approximately every 6 milliseconds.

\subsubsection{Trigger Derived Timestamp}\label{trigger_deri_time}

The finest resolution timestamp is derived from the raw HAWC TDC data itself. 
The raw HAWC TDC data have a field called `Extended Trigger Time Tag' which contains the trigger arrival time ($t_0$) relative to the last bunch counter reset (CRST) with a precision of 800 ns. 
For each of the 28 TDC channels which corresponds to the TDC timestamp, there is a rising edge time measurement, again relative to the last TDC reset, lets call it $t_{i}^{+}$ where $i=1,...,28$. 
Now each of the 28 channels will provide a delta time measurement ($\Delta t_i$) from the most recent rising edge of its input signal until the arrival of the Trigger signal.
\\
\begin{equation}\label{eq01}
\Delta t_i = t_0 - t_{i}^{+}
\end{equation}
\\
Thus the finest resolution time is given by,
\\
\begin{equation}\label{eq02}
\Delta t = \frac{1}{28} \sum_{i=1}^{28} \Delta t_i.
\end{equation}
\\
The $\Delta t$ measured in this method has an accuracy of 0.1 ns(100 ps). 

In the final step, we construct the GPS timestamp of the trigger by combining these three components:
\begin{center}
GPS Time = NTP Time + TDC Time + $\Delta t$.
\end{center}
We thus have an accurate timestamp for the arrival time of the Trigger signal.
The absolute accuracy of this time stamp is 1 $\mu$s, which is the absolute accuracy of the TDC time, 
note that following a similar procedure we can measure all of the PMT signal edges with respect to the arrival of the Trigger signal.

\section{Control System}

The Control System provides several services to HAWC:
1) keep all TDCs working synchronously,
2) issue a synchronous trigger signal to the TDCs,
3) issue a scaler DAQ trigger signal called Load Next Event (LNE) and send the status of the detector to the scaler system.
Other than these major services, the Control system also has a general purpose level shifter to shift signals from LVDS to ECL and vice versa.

\subsection{Hardware implementation.}

\begin{figure}[h!]
\centering
\includegraphics[width=100mm]{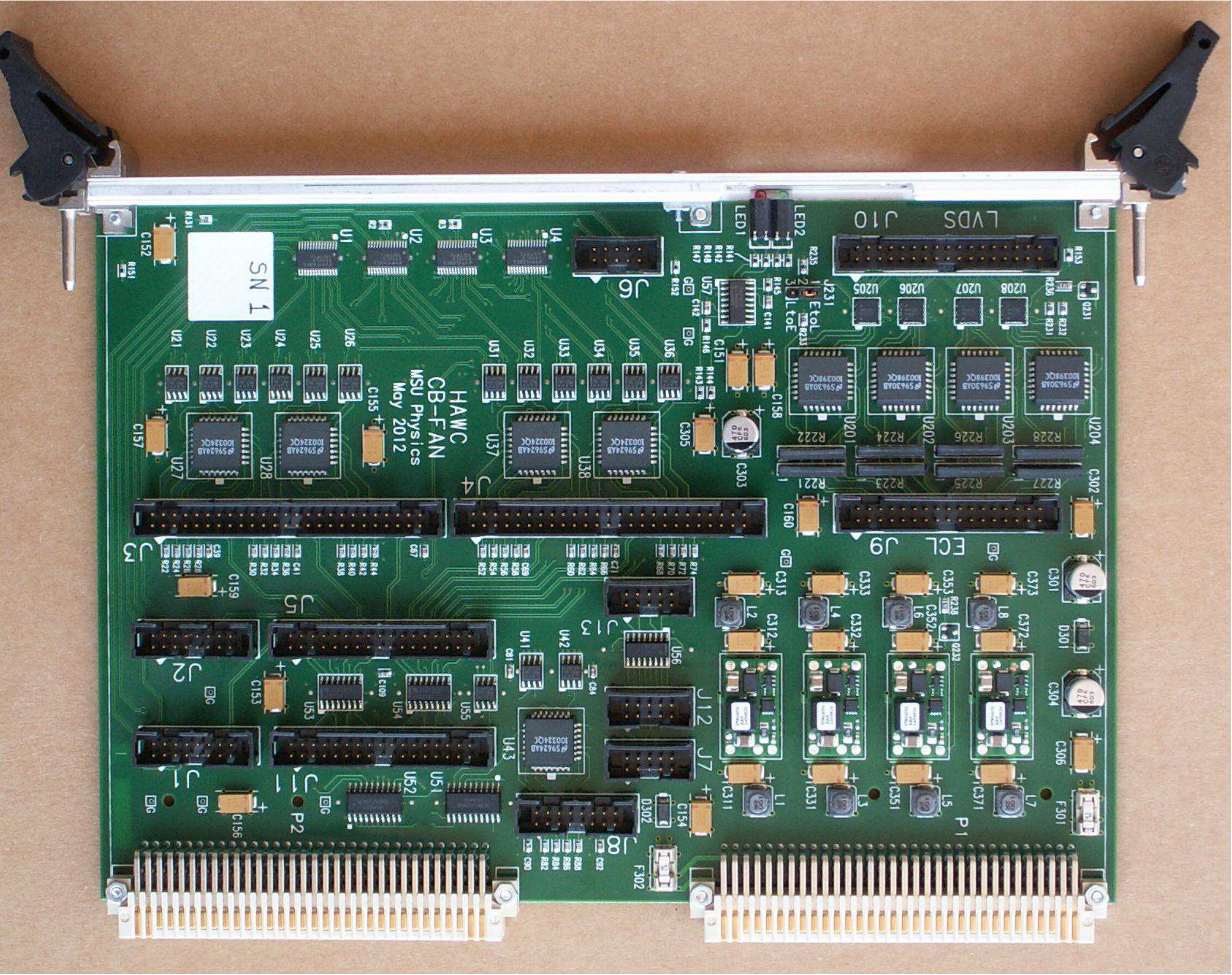}
\caption{A photograph of the fully assembled CB\_Fan card is shown.
        A CB\_Fan card is able to provide the level conversions and Fan-outs required to handle 6 TDCs.
}\label{CBFanCard}
\end{figure}

The Control System is made from two custom VME boards: a Control type HClock Card and a CB\_Fan Card.
The Control type HClock Card is a version of the HClock card with 6 input ports and 4 output ports.
The Control type HClock Card also gets the 40 MHz global clock from the Clock card.
The CB\_Fan card is a 2 slots wide 6U VME-64X module, that is designed to provide appropriate level conversions and fan-outs for the Control type HClock Card-TDC interface.
The CB\_Fan card does not perform any logic.
A photograph of a fully assembled CB\_Fan card is shown in Figure~\ref{CBFanCard}.\\

\begin{figure}[h!]
\centering
\includegraphics[width=100mm]{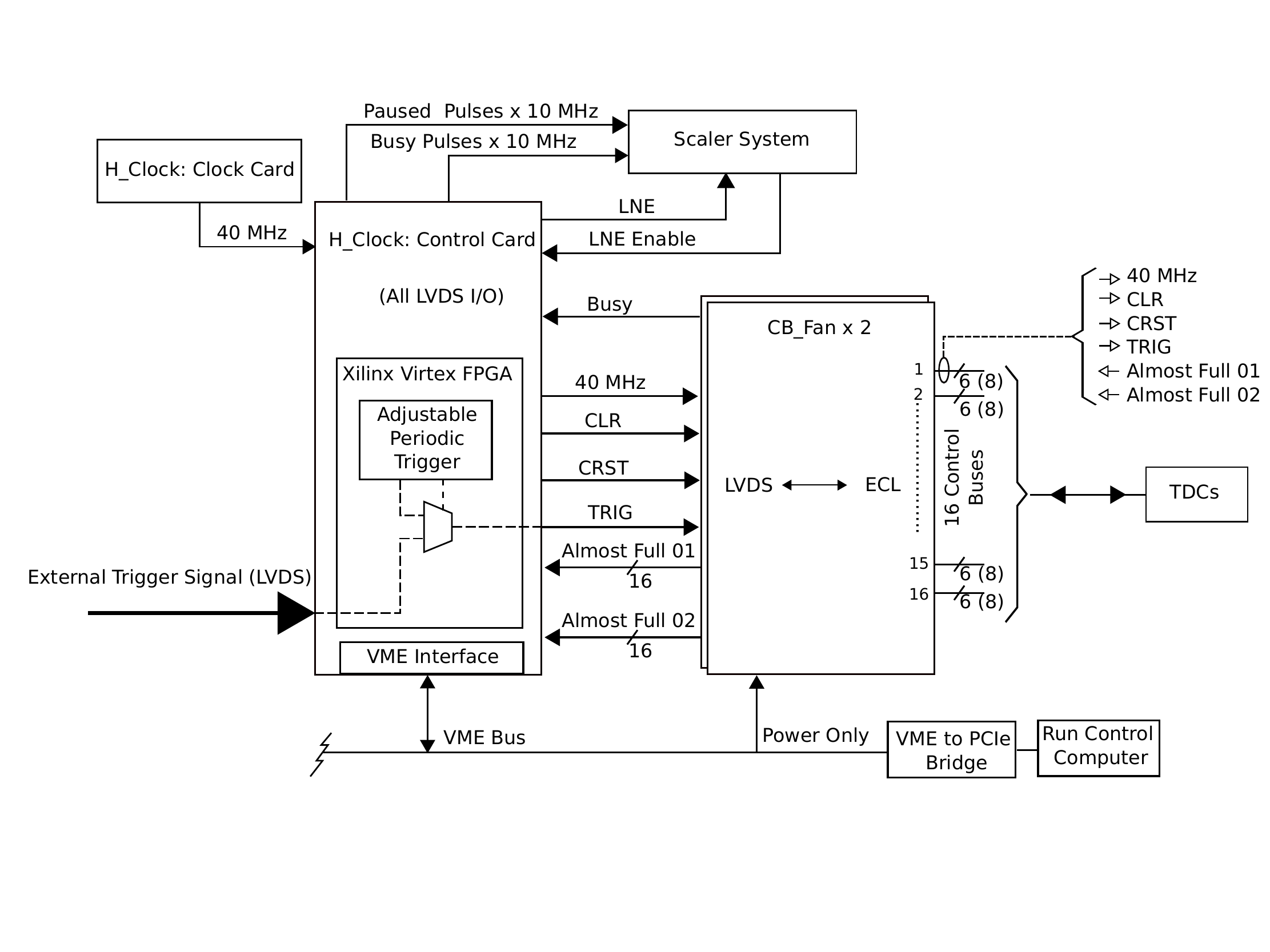}
\caption{A block diagram of the Control System is shown.
         The Control system consists of two VME cards: Control Card and CB\_Fan card.
         The Control card does all the logical operations and the CB\_Fan card does the appropriate level conversion to interface TDCs to the Control Card.}\label{ControlCardSchem}
\end{figure}

A schematic diagram of the connections between the Clock type HClock Card, Control type HClock Card, CB\_Fan card and the scaler system is shown in Figure~\ref{ControlCardSchem}.
The input signal coming from the clock Card is the 40 MHz global clock signal.
The Control type HClock Card makes several copies of this 40 MHz clock signal and distributes it to the $17^{th}$ signal pair of all the GPIO connectors and to the FPGA.
The interface between the Control type HClock Card and the Scaler system consists of three outputs and one input: 10 MHz reference, Pause Pulses, Busy Pulses, LNE (the trigger signal for the scaler system) and LNE Enable input.
The 10 MHz reference is a continuous 10 MHz square wave signal output.
The Pause Pulses produces a 10 MHz signal in-phase with the 10 MHz reference when the Control system is in the pause state.
The scaler system counts both of these signals.
The ratio of the number of pause pulses to the number of 10 MHz reference pulses gives the 
fractional dead time of the detector enforced by the experiment control system.
The functionality of the Busy Pulse output is similar to the Pause Pulses except Busy Pulses produce a 10 MHz signal 
when at least one TDC is filled to the almost full level.
The Load Next Event (LNE) signal, a 100 Hz clock, acts as a readout-start trigger for the scaler system.\\

The Control type HClock Card interface to a CB\_Fan card consists of four output signals, 40 MHz, CLR, CRST and TRIG, and 16 input signals, 16 Almost Full.
The four output signals are the TDC control-bus signals.
The Control card makes four identical copies of all output signals and can accept up to 64 inputs.
Therefore, one Control type HClock Card can be connected with four CB\_Fan cards.

But the TDCs can not directly connect to the Control type HClock Card, because these I/Os are LVDS signals but the TDC control bus is compatible with only ECL signals.
The CB\_Fan card is designed to provide level shifting between the HClock Card LVDS signals and TDC ECL signals.
Apart from the level shifting, the CB\_Fan card makes 6 identical copies of the 40 MHz, CLR, CRST and TRIG signals.
The active edges of the TRG, CLR and CRST signals are placed midway between the active CLOCK edges.
Control Firmware, Control type HClock card hardware and CB\_Fan card hardware is designed obtain these signals synchronized to better than 12.5 ns.

Therefore, one CB\_Fan card can be used to control up to 6 TDCs.
Since one Control type HClock Card can interface with 4 CB\_Fan cards, the GTC system is capable of controlling up to 24 TDCs.\\

\subsection{Firmware implementation.}
\label{sec:ControlFirmware}

\begin{figure}[h!]
\centering
\includegraphics[width=100mm]{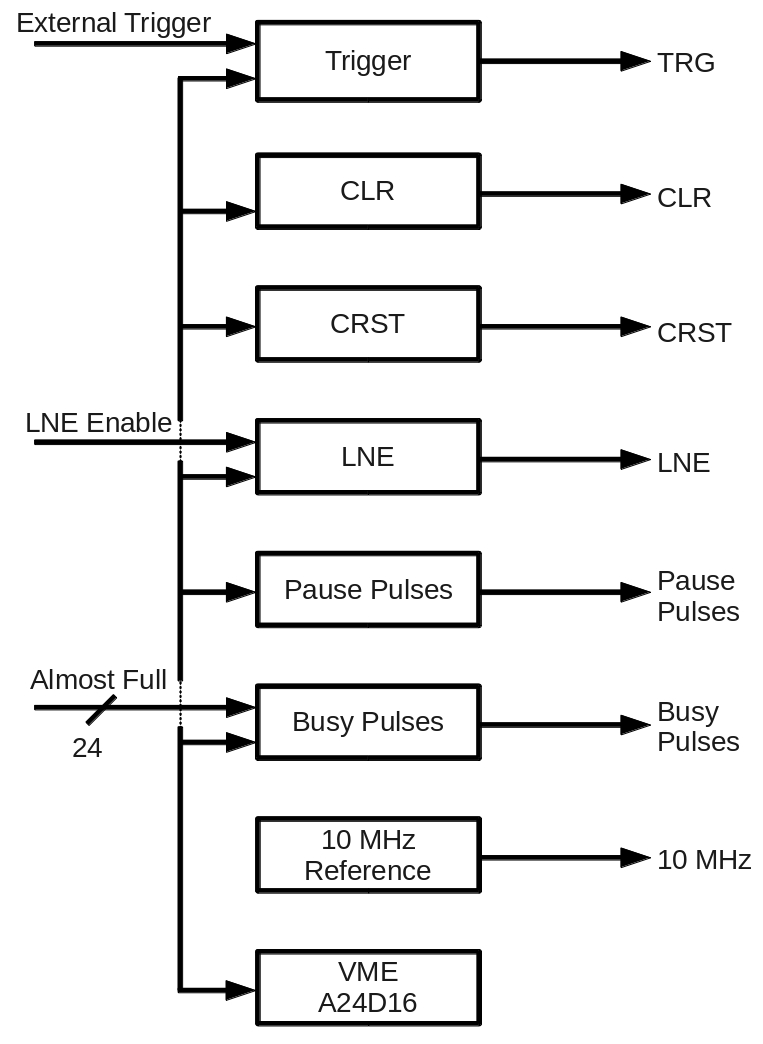}
\caption{A simplified functional block diagram of the Control firmware is shown.
        The Trigger, CLR, and CRST modules generate the TDC control signals, and LNE, Pause Pulses, Busy Pulses and 10 MHz Reference produces signals to the scaler system.
        Similar to the Control firmware, communication between the Control system and the control computer was done by an A24D16 VME interface.
}\label{Fig:ControlFirmware}
\end{figure}

A simplified functional block diagram of the Control firmware is shown in Figure~\ref{Fig:ControlFirmware}.
Similar to the Clock firmware, this firmware is also a sequential logic design implemented in VHDL.
However, unlike the Clock firmware individual modules of the Control firmware are not connected in series.
Coordination of these modules is done through the VME module.\\

The first module shown in Figure~\ref{Fig:ControlFirmware} is the trigger module, which coordinates the trigger signals that go to the TDCs.
The trigger module can work in three modes: pause, periodic trigger, and external trigger.
In the pause mode, the trigger module does not issue any triggers.
In the periodic trigger mode, the trigger module issues a periodic trigger signal with a known frequency set by the VME module.
In the external trigger mode, the trigger module issues a trigger signal upon a request coming from the external trigger.
This trigger mode is not currently used in HAWC; the potential usage of this functionality is discussed in section~\ref{sec:pottentialusage}.
In a typical data taking run of HAWC, the trigger module runs in the periodic trigger mode with a trigger frequency of 40 kHz.
At the end of each run the HAWC experiment control system sends a request via the GTC control software to the VME module to switch the trigger module to the pause mode.
The 40 kHz periodic trigger frequency was chosen because it is the optimum trigger frequency for the HAWC DAQ system\footnote{More details about the HAWC DAQ will be discussed in a future paper in preparation.}.
However, this periodic trigger frequency can be changed by a request to the VME module from the HAWC run control system.
The CLR and CRST modules issue the clear and reset signals to the TDCs upon a request coming from the VME module.
These requests originate in the run control system at the beginning of each run.\\

The next three modules in Figure~\ref{Fig:ControlFirmware} provide the signals to the scaler system.
The 10 MHz reference module is a 10 MHz square wave signal generator that generates a reference pulse stream to the scaler system.
The functionality of the Pause Pulse module is equivalent to a multiplexer with two inputs and one output: Logic Lo input, 10 MHz square wave input and Pause Pulse output.
When the Trigger module is in the Pause state, the Pause Pulse output switches to the 10 MHz square wave.
When the Trigger module is not in the Pause state, the output switches to the Logic Lo level.
The Busy Pulse module has a similar functionality, except that the selection between Logic Lo and 10 MHz square is done using the OR of the Almost Full signals.
If any of the Almost Full inputs are Logic Hi, the Busy Pulse output gets connected with the 10 MHz square wave, otherwise the output stays in the Logic Lo level.
Therefore, one can calculate the fraction of the time that HAWC stays in the busy state using the ratio between Busy Pulses per run and 10 MHz square wave pulses.

\section{Additional Capabilities}
\label{sec:pottentialusage}
Apart from the main features of the GTC system described above it is also able to support several other functionalities:
 external triggers, SBC readout signal and LVDS control busses.
At the present, HAWC runs using a periodic trigger signal.
However, the GTC system is designed to support both the periodic trigger mode and the external trigger mode.\\

The SBC read out signal is another currently unused feature of the GTC system.
Similar to the other signals this signal also comes from the Control card and goes to the CB\_Fan card.
The CB\_Fan card converts this signal to a single ended 3.3V logic level 110 Ohm back terminated signal and makes 8 copies of them.
The intention of this signal is to issue a read out request to the SBCs.
One of the potential uses of this signal is to issue an SBC read out request when at least one TDC becomes almost full.\\

Besides the ECL signal TDC control buses, each CB\_Fan card also fans-out three copies of LVDS control buses.
This control bus signaling could drive the read out of additional devices such 
as PMT digitizers or external programed trigger modules.

\section{Discussion}
In order to check reliability of the GTC system, we made a test setup with two independent GTC systems and a 
TDC with an accuracy of 200 ps to record timestamps coming from both TDCs.
If both GTC systems are reliable we expect them to produce identical timestamps for a given trigger.
We collected data during 24 hours and didn't print any event with unequal timestamps.\\




The health of the GTC system was continuously monitored from early 2013, and we found that the 1PPS signal had a jitter, less than 50 ns, with respect to the 10 MHz output.
Each time the 1PPS moved more than 25 ns it caused an overwrite of the Clock firmware's internal clock from the GPS clock.
Therefore, this jitter produced error flags in the timestamps.
The average rate of this error flag is 11 per hour.
This jitter introduced an upper limit of 25 ns accuracy to the GTC generated timestamp.
However, the 25 ns accuracy is well below the required accuracy of 1$\mu$s for HAWC.
NAVSYNCH builds a new GPS module with an internal phase lock to lock the 1PPS and 10 MHz outputs.
However, the present GPS module is sufficient for HAWC's requirements.\\

\section{Summary}
The HAWC gamma ray observatory equipped with a fully functional GTC system started its first phase, with 30 tanks, in November of 2012.
The PMT signals were digitized using a Caen VX1190A TDC.
Apart from the PMT signals the Clock system generates a 32 bit timestamp encoded in a 32 channel pulse pattern, which is similar to the TOT signals of the PMT output signals after the FEBs.
These 32 signals were digitized using another Caen VX1190A TDC.
Both TDCs were read out by their own SBCs via a VME back plane and the data is transferred to the online reconstruction farm via an Ethernet connection.
In the online reconstruction farm the timestamp and the PMT data that correspond to the same Event IDs are combined to form a single event.
After combining these two parts, the online reconstruction software decodes the timestamp.\\

In order to make TDCs work synchronously, the Control system delivers two identical copies of a 40 MHz clock signal and the trigger signal to TDCs.
Since the online reconstruction process uses the Event ID to combine the PMT data with the timestamp, it is a must to maintain a unique Event ID to event fragments that correspond to a given trigger.
Therefore, at the beginning of each run the GTC system issues a clear (CLR) signal to reset the Event ID counters. 
The GTC system also issues a reset (CRST) signal at the beginning of each run to reset all the other counters in TDCs.\\

The health monitoring of the GTC system was continuously done from early 2013 and it reveals that the accuracy of the timestamps produced by the GTC system has an upper limit of 25 ns.
However, the 25 ns accuracy is well below the required accuracy of 1$\mu$s for HAWC.\\

When HAWC is completed in 2015, it will have 300 tanks instrumented with 4 PMTs per each tank.
This will increase the number of TDCs required to record PMT signals up to 10, and another TDC will be used to record timestamps.
Therefore, the completed HAWC will be instrumented with 11 TDCs.
The GTC system with two CB\_Fan cards will be able to match this requirement.

\section{Acknowledgment}
We would like to give our special thanks to everyone in the HAWC collaboration who helped us 
to design and build the GTC system.
Funding for the GTC system construction was provided by the NSF HAWC construction grant, PHY 1002546, via a subcontract with the University of Maryland, and NSF HAWC grants PHY 0901973 and PHY 1002432.

\bibliographystyle{elsarticle-harv}



\end{document}